\documentclass[onecolumn,showpacs]{article}
\usepackage{adjustbox}
\usepackage{graphicx}
\usepackage{dcolumn}
\usepackage{bm}
\usepackage{amsmath}
\usepackage{subcaption}
\usepackage{amssymb}
\usepackage{multirow}
\usepackage{caption}
\usepackage{epstopdf}
\usepackage{hyperref}
\usepackage{wrapfig}
\usepackage{lipsum} 

\begin{document}
{\setlength{\oddsidemargin}{1.2in}
\setlength{\evensidemargin}{1.2in} } \baselineskip 0.55cm
\begin{center}
{\LARGE {\bf Phase space analysis of cosmological models on $f(R,G,T)$ gravity}}
\end{center}
\date{\today}
\begin{center}
  Elangbam Chingkheinganba Meetei, S. Surendra Singh \\
   Department of Mathematics, National Institute of Technology Manipur,\\ Imphal-795004,India\\
   Email:{ chingelang@gmail.com, ssuren.mu@gmail.com }\\
\end{center}
\begin{center}
 \textbf{Abstract}
 \end{center}
 In the present article, we developed a dynamical system in the context of modified $f(R,G,T)$ gravity, where $R$, $G$ and $T$ are Ricci scalar, Gauss-Bonnet term and energy-momentum tensor respectively. Development of the dynamical system is done by first defining 9 dimensionless variables and formulate a ordinary differential equations by taking derivative of the variables with respect to logarithmic time $N=\log a(t)$, where $a(t)$ is the scale factor. This formulation is applied to the model $f(R,G,T)=\alpha R^{l}+\beta G^{m} + \gamma T^{n} $, and its solutions and their corresponding stabilities are analysed in details. From the plot of density parameters vs $N=-log(1+z)$, we conclude that our Universe is currently dominated by dark energy, which is compatible with current observation. Taking $l=2$, cosmic parameters such as deceleration parameter $(q)$, equation of state $(\omega)$ and state finder parameters are also discussed by fixing one dimensionless variable, showing our Universe's expansion is accelerating with the present value of $q=-0.5$. Present value of $\omega=-0.66$ suggests that our Universe is in Quintessence phase. Lastly, the validity of the model with respect to the $\Lambda$CDM model is checked by using 77 Hubble, 15 BAO and 1024 pantheon datasets, which implies that our model is aligned with the $\Lambda$CDM model's bahaviour. \\

\section{Introduction}\label{sec1}
The cosmos has undergone a remarkable transformation since its inception approximately $13.8$ billion years ago with the Big Bang, which marked the origin of space, time, and matter. Following a brief period of rapid inflation, the Universe cooled enough to allow the formation of light elements through nucleosynthesis. As it expanded further, atoms formed, leading to the decoupling of light and the release of the cosmic microwave background radiation. The subsequent “dark ages” ended with the formation of the first stars and galaxies, triggering the epoch of re-ionization. Over cosmic time, gravitational interactions drove the large-scale structuring of the Universe, the synthesis of heavier elements in stars, and the evolution of galaxies. In the more recent cosmic past, observations reveal an accelerating expansion attributed to dark energy, shaping the current state of the Universe and driving ongoing exploration in cosmology \cite{A.G. Riess}. Dark Energy (DE) stands out as one of the most intriguing discoveries in contemporary cosmology and was first explored through observations of Type Ia Supernovae (SNe Ia) \cite{S. Perlmutter}. Researchers have investigated DE using various approaches, such as observational Hubble parameter measurements, Cosmic Microwave Background Radiation (CMBR), power spectrum analyses, and the large-scale structure of the Universe \cite{A.G. Riess,S. Perlmutter,S. Perlmutter M.S.,D.J. Eisenstein}. Current estimates indicate that the Universe consists of roughly $74\%$ dark energy, $22\%$ dark matter, and $4\%$ ordinary matter.

The Friedmann-Lemaître-Robertson-Walker (FLRW) framework models the Universe's large-scale structure as homogeneous and isotropic across four dimensions. When combined with cold dark matter (CDM), the $\Lambda$CDM model forms a highly effective cosmological paradigm for analyzing galactic behavior \cite{L. Baudis}, surpassing the explanatory power of the cosmological constant $\Lambda$ alone \cite{S. Weinberg}. Incorporating an early phase of inflation, the $\Lambda$CDM model aligns well with a wide range of astronomical observations across different cosmic scales \cite{L. Pérénon}. However, its validity has come under scrutiny in recent years. Originally formulated to match Hubble parameter data, the model now faces the "Hubble tension"---a significant disparity between the Hubble constant ($H_0$) inferred from early-Universe predictions \cite{A.G. Riess} and values obtained from direct, model-independent late-time observations \cite{Planck2018}. Furthermore, the $\Lambda$CDM framework also suffers from fine-tuning issues that challenge its theoretical robustness \cite{S. M. Carroll}.

In 2005, S. Carloni et al. analyse the evolution in $f(R)=R^{n}$ models \cite{Carloni}. In 2008, Amendola, Tsujikawa introduced phae-space analysis of $f(R)$ models and classified cosmological dynamics \cite{amendola}. In 2009, De Felice, Tsujikawa studied stability and late-time attractors of $f(G)$ models \cite{Felice}. In 2012, Zhang, Li, Gong, et al. presented fixed point analysis of specific $f(T)$ forms \cite{zang}. In 2018, S. Santos da Costa et al. presented dynamical analysis on $f(R,G)$ cosmology \cite{Santos}. In 2024, Amit Samaddar and Surendra presented dynamical system approach of interacting dark energy models in $f(R,T^{\phi})$ gravity \cite{Surrendra A.}.  Some authors have also presented dynamical system analysis on different gravity to study the early time inflation and late-acceleration of the Universe and also the evolution of dark energy \cite{Sonia S.,Shivangi S.,s. SS}. 

The dynamical system of modified gravity refers to the formulation of gravitational theories-beyond Einstein's General Relativity (GR)—in the language of dynamical systems theory, which is a mathematical framework used to analyze the behavior of systems that evolve over time. This approach is particularly useful in cosmology, where one studies the evolution of the Universe by turning the cosmological field equations (usually differential equations derived from the modified gravity theory) into an autonomous dynamical system and analyzing its fixed points and stability. These points can represent cosmological phases like inflation, matter domination, or late-time acceleration (dark energy-like behavior). 

In the present paper, we analyse the $f(R,G,T)$ gravity that transforms the general theory of relativity action by the $R$, $G$ and $T$. The paper is divided into sections as: In section \ref{sec2}, $f(R,G,T)$ gravity and its cosmology are discussed. In section \ref{sec3}, an autonomous dynamical system in the context of a specific model, $f(R,G,T)=\alpha R^{l}+\beta G^{m} + \gamma T^{n} $, is developed. In section \ref{sec4}, critical points and their stability are discussed. In section \ref{cosmic-p}, cosmic parameters are discussed. In section \ref{comparison}, we checked the validity of the model by using observational data with respect to the standard $\Lambda$CDM model. Lastly, section \ref{conculsion} presents the result and discussion of the paper.

\section{$f(R,G,T)$ gravity and its cosmology}\label{sec2}
The action for this gravity is given by \cite{U. Debnath}
  \begin{equation}\label{action}
    S_{f(R,G,T)} = \frac{1}{2\kappa^{2}}\int d^{4}x \sqrt{-g} f(R, G, T) + \int d^{4}x \sqrt{-g} (\mathcal{L}_{m}+\mathcal{L}_{r})
  \end{equation}
  where $ \kappa^{2}=8\pi G $, $G$ is Newtonian constant, $ \mathcal{L}_{m} $ and $\mathcal{L}_{r}$ denote the matter Lagrangian and radiation lagrangian respectively. $R$ denotes the Ricci scalar, $G$ denotes the Gauss Bonnet Invariant and $g$ denotes the metric determinant of the fundamental metric $g_{\mu\nu}$.
  The Gauss Bonnet Invariant is defined by
\begin{equation}\label{equation 2}
G \equiv R^{2}-4R^{\mu\nu}R_{\mu\nu}+R^{\mu\nu\alpha\beta}R_{\mu\nu\alpha\beta}
\end{equation}
Here, $R_{\mu\nu}$ is the Ricci tensor and $R_{\mu\nu\alpha\beta}$ is the Riemann tensor.
The energy momentum tensor $T_{\mu\nu}$ is defined as
\begin{equation}\label{equation 3}
  T_{\mu\nu}=-\frac{2}{\sqrt{-g}}\frac{\partial (\sqrt{-g}\mathcal{L}_{m})}{\partial g^{\mu\nu}}
\end{equation}

Taking variation of the action (\ref{action}) w.r.t the fundamental metric $g_{\mu\nu}$ and after required manipulation the gravitational field equation is obtained as follow
\begin{equation}\label{equation 4}
\begin{split}
\kappa^{2} T^{m}_{\mu\nu} &= R_{\mu\nu} f_{R} - \frac{1}{2} g_{\mu\nu} f - (g_{\mu\nu} \Box - \nabla_{\mu} \nabla_{\nu}) f_{R} + (2 R R_{\mu\nu} - 4 R^{\lambda}_{\nu} R_{\lambda\mu} + 2 R_{\mu}^{\lambda \kappa \alpha} R_{\nu \lambda \kappa \alpha} \\
&\quad - 4 g^{\lambda \alpha} g^{\kappa \beta} R_{\mu \lambda \nu \kappa} R_{\alpha \beta}) f_{G} + 2 R g_{\mu\nu} \Box f_{G} - 2 R \nabla_{\mu} \nabla_{\nu} f_{G} + 4 R^{\lambda}_{\mu} \nabla_{\lambda} \nabla_{\nu} f_{G}-4R^{\mu\nu}\Box f_{G} \\
&\quad -4g_{\mu\nu}R^{\lambda\kappa}\nabla_{\lambda}\nabla_{\kappa}f_{G}+4R^{\lambda}_{\nu}\nabla_{\lambda}\nabla_{\mu}f_{G}+4g^{\lambda\alpha}g^{\kappa\beta}R_{\mu\lambda\nu\kappa}\nabla_{\alpha}\nabla_{\beta}f_{G} \\ &\quad +f_{T}(\theta_{\mu\nu}+T_{\mu\nu}),
\end{split}
\end{equation}
where $f\equiv f(R, G ,T)$, $f_{R}=\frac{\partial f}{\partial R}$, $f_{G}=\frac{\partial f}{\partial G}$, $f_{T}=\frac{\partial f}{\partial T}$, $\theta_{\mu\nu}\equiv g^{\alpha\beta}\frac{\partial T_{\alpha\beta}}{\partial g^{\mu\nu}}$, $\Box\equiv g^{\mu\nu}\nabla_{\mu}\nabla_{\nu}$ and $\nabla_{\mu}$ represents the covariant derivative w.r.t. $x_{\mu}$.

   In this analysis, the FLRW metric of a isotropic and flat Universe is examined. Its mathematical expression in cartesian coordinates, is given by

  \begin{equation}\label{metric}
ds^{2}=-dt^{2}+a^{2}(t)(dx^{2}+dy^{2}+dz^{2}),
\end{equation}
where $a(t)$ is the scale factor and $t$ is the cosmic time. The Hubble parameter $H = \frac{\dot{a}}{a}$, describes the rate of expansion of the Universe. The Gauss Bonnet Invariant and the Ricci scalar are given by
\begin{equation}\label{ricci}
R=6(\dot{H}+2H^{2}),
\end{equation}
\begin{equation}\label{gauss}
G=24H^{2}(\dot{H}+H^{2}).
\end{equation}

Using the action (\ref{action}) and the metric (\ref{metric}), the field equations are given by
\begin{equation}\label{first}
 3H^{2}f_{R}=\kappa^{2}(\rho_{m}+\rho_{r})+\frac{1}{2}Gf_{G}-\frac{1}{2}f+\frac{1}{2}R{f_{R}}-3H\dot{f_{R}}-12H^{3}\dot{f_{G}}+f_{T}(\rho_{m}+\frac{4}{3}\rho_{r})
\end{equation}
and
\begin{equation}\label{second}
  2f_{R}\dot{H}=-\kappa^{2}(\rho_{m}+\frac{4}{3}\rho_{r})-\ddot{f_{R}}+H\dot{f_{R}}-4H^{2}\ddot{f_{G}}-4H(2\dot{H}-H^{2})\dot{f_{G}}.
\end{equation}

Here, $\rho_{m}$ and $p_{m}$ are the energy density and pressure of the matter respectively. The corresponding basic conservation equations are given by
\begin{equation}\label{rho-m}
  \dot{\rho_{m}}+3H\rho_{m}=0,\quad p_{m}=0.
\end{equation}
\begin{equation}\label{rho-r}
  \dot{\rho_{r}}+4H\rho_{r}=0,\quad p_{r}=\frac{1}{3}\rho_{r}.
\end{equation}
 
 Equations (\ref{first}) and (\ref{second}) are rewritten by providing new definitions as
 \begin{equation}\label{1}
   3H^{2}f_{R}=\kappa^{2}(\rho_{m}+\rho_{r}+\rho_{de})
 \end{equation}
 \begin{equation}\label{1}
    2f_{R}\dot{H}=-\kappa^{2}(\rho_{m}+\frac{4}{3}\rho_{r}+p_{de}+\rho_{de})
 \end{equation}
 where $\rho_{de}$ and $p_{de}$ are defined as 
 \begin{equation}\label{rhoDE}
   \kappa^{2}\rho_{de}=\frac{1}{2}Gf_{G}-\frac{1}{2}f+\frac{1}{2}R{f_{R}}-3H\dot{f_{R}}-12H^{3}\dot{f_{G}}+f_{T}(\rho_{m}+\frac{4}{3}\rho_{r})
 \end{equation}
 and
 \begin{equation}\label{pDE}
   -\kappa^{2}(\rho_{de}+p_{de})=-\ddot{f_{R}}+H\dot{f_{R}}-4H^{2}\ddot{f_{G}}-4H(2\dot{H}-H^{2})\dot{f_{G}}.
 \end{equation}
 The equation of states for the dark energy is defined by
 \begin{equation}\label{DEeos}
   \omega_{de}=-1-\frac{-\ddot{f_{R}}+H\dot{f_{R}}-4H^{2}\ddot{f_{G}}-4H(2\dot{H}-H^{2})\dot{f_{G}}}{\frac{1}{2}Gf_{G}-\frac{1}{2}f+\frac{1}{2}R{f_{R}}-3H\dot{f_{R}}-12H^{3}\dot{f_{G}}+f_{T}(\rho_{m}+\frac{4}{3}\rho_{r})}
 \end{equation} 
  and that of the effective equation of state is defined as
  \begin{equation}\label{eff}
    \omega_{eff}=-1-\frac{2\dot{H}}{3H^{2}}
  \end{equation}
   
   Also density parameters for the matter, radiation and dark energy are defined by 
   \begin{equation}\label{density}
     \Omega_{m}=\frac{\rho_{m}}{3H^{2}f_{R}},\quad \Omega_{r}=\frac{\rho_{r}}{3H^{2}f_{R}},\quad\Omega_{de}=\frac{\frac{Gf_{G}}{2}-\frac{f}{2}+\frac{Rf_{R}}{2}-3H\dot{f_{R}}-12H^{3}\dot{f_{G}}+f_{T}(\rho_{m}+\frac{4}{3}\rho_{r})}{3H^{2}f_{R}}
   \end{equation}

Now, using equations (\ref{rho-m}) and (\ref{rho-r}), equation (\ref{first}) can be written as follow
\begin{equation}\label{15}
  1=\frac{\rho_{m}}{3H^{2}f_{R}}+\frac{\rho_{r}}{3H^{2}f_{R}}+\frac{Gf_{G}}{6H^{2}f_{R}}-\frac{f}{6H^{2}f_{R}}+\frac{R}{6H^{2}}-\frac{\dot{f_{R}}}{Hf_{R}}-\frac{4H\dot{f_{G}}}{f_{R}}+\frac{f_{T}\rho_{m}}{3H^{2}f_{R}}+\frac{4f_{T}\rho_{r}}{9H^{2}f_{R}}
\end{equation}

\section{Construction of dynamical systems}\label{sec3}
A dynamical system in cosmology refers to the use of mathematical techniques from dynamical systems analysis to study the evolution of the Universe based on cosmological models, particularly within the framework of general relativity. Typically FLRW metric, which assumes a homogeneous and isotropic Universe, is used to set up dynamical system. By redefining variables and re-scaling time i.e. $N=log(a(t))$, the cosmological equations can often be rewritten as a system of autonomous differential equations: $\frac{dx}{dN}=f(x)$, $x$ is a vector in $R^{n}$ space. The study of dynamical system helps in understanding cosmic inflation, analyzing dark energy models, exploring modified gravity theories and studying early and late-time behaviour of the Universe.
 
 In the present section, dynamical systems of a specific model of the extended $f(R,G,T)$ models viz. $\alpha R^{l}+\beta G^{m} +\gamma T^{n}$, where $l$, $m$, $n$, $\alpha$, $\beta$ and $\gamma$ are free non-zero positive parameters, are developed using equations (\ref{first}) and (\ref{second}). Let us introduce the dimensionless variables as given below:

$d_{1}=\frac{\rho_{m}}{3H^{2}f_{R}}$,\quad$d_{2}=\frac{\rho_{r}}{3H^{2}f_{R}}$,\quad$u=\frac{Gf_{G}}{6H^{2}f_{R}}$, \quad$v=\frac{f}{6H^{2}f_{R}}$,\quad$w=\frac{R}{6H^{2}}$,\quad $x=\frac{\dot{f_{R}}}{Hf_{R}}$,\quad$y=\frac{4H\dot{f_{G}}}{f_{R}}$, \\
$z_{1}=\frac{f_{T}\rho_{m}}{3H^{2}f_{R}}$,\quad$z_{2}=\frac{f_{T}\rho_{r}}{3H^{2}f_{R}}$.

 Using a dimensionless time variable $N=\log a(t)$, the following dynamical system is obtained by the derivative of these variables with respect to $N$. For our present framework of gravity, the dynamical system is given by:
 \begin{equation}\label{DS}
   \begin{split}
      \frac{dd_{1}}{dN}= & -3d_{1}-2d_{1}\frac{\dot{H}}{H^{2}}-d_{1}x \\
      \frac{dd_{2}}{dN}= & -4d_{2}-2d_{2}\frac{\dot{H}}{H^{2}}-d_{2}x \\
      \frac{du}{dN}=  & y\frac{G}{24H^{4}}+u\frac{\dot{G}}{GH}-2u\frac{\dot{H}}{H^{2}}-ux \\
       \frac{dv}{dN}= & \frac{\dot{f}}{6H^{3}f_{R}}-2v\frac{\dot{H}}{H^{2}}-vx \\
       \frac{dw}{dN}= & \frac{\dot{R}}{6H^{3}}-2w\frac{\dot{H}}{H^{2}} \\
      \frac{dx}{dN}=  & \frac{\ddot{f_{R}}}{H^{2}f_{R}}-x\frac{\dot{H}}{H^{2}}-x^{2} \\
       \frac{dy}{dN}= & \frac{4\ddot{f_{G}}}{f_{R}}+y\frac{\dot{H}}{H^{2}}-yx \\
       \frac{dz_{1}}{dN}= & d_{1}\frac{\dot{f_{T}}}{H}-3z_{1}-2z_{1}\frac{\dot{H}}{H^{2}}-z_{1}x\\
       \frac{dz_{2}}{dN}= & d_{2}\frac{\dot{f_{T}}}{H}-4z_{2}-2z_{2}\frac{\dot{H}}{H^{2}}-z_{2}x
   \end{split}
 \end{equation}
  To be a closed system, the right hand side of the above equations must be expressed in terms of the variables defined above. The non variable factors or terms of the dynamical system are given by:
\begin{equation}\label{26}
\begin{split}
  \frac{\dot{H}}{H^{2}}&= w-2\\
  \frac{G}{24H^{4}}&= w-1\\
  \frac{\dot{R}}{6H^{3}}&= \frac{wx}{a}\\
  \frac{\dot{G}}{HG}&= \frac{1}{w-1}\bigg(\frac{wx}{a}+2(w-2)^{2}\bigg)\\
  \frac{\dot{f}}{6H^{3}f_{R}}&= \frac{wx}{a}+\frac{u}{w-1}\bigg(\frac{wx}{a}+2(w-2)^{2}+\frac{3}{2}z_{1}\bigg)\\
  \frac{4\ddot{f_{G}}}{f_{R}}&=-3d_{1}-4d_{2}+x-2(w-2)+y(5-2w)-\eta\\
  d_{1}\frac{\dot{f_{T}}}{H}&=-3bz_{1}\\
   d_{2}\frac{\dot{f_{T}}}{H}&=-3bz_{2}
  \end{split}
\end{equation}  
where $a=\frac{dlnf_{R}}{dlnR}$, $b=\frac{dlnf_{T}}{dlnT}$, and $\eta=\frac{\ddot{f_{R}}}{H^{2}f_{R}}$.
Also, by using equation (\ref{15}), we have
\begin{equation}\label{27}
  1=d_{1}+d_{2}+u-v+w-x-y+z_{1}+\frac{4}{3}z_{2}.
\end{equation}
Now the dynamical system (\ref{DS}) becomes
\begin{equation}\label{DS1}
   \begin{split}
      \frac{dd_{1}}{dN}= & -3d_{1}-2d_{1}(w-2)-d_{1}x \\
      \frac{dd_{2}}{dN}= & -4d_{2}-2d_{2}(w-2)-d_{2}x \\
      \frac{du}{dN}=  & y(w-1)+\frac{u}{w-1}\big(\frac{wx}{a}+2(w-2)^{2}\big)-2u(w-2)-ux \\
       \frac{dv}{dN}= & \frac{wx}{a}+\frac{u}{w-1}\big(\frac{wx}{a}+2(w-2)^{2}\big)+\frac{3}{2}z-2v(w-2)-vx \\
       \frac{dw}{dN}= & \frac{wx}{a}-2w(w-2) \\
      \frac{dx}{dN}=  & \eta-x(w-2)-x^{2} \\
       \frac{dy}{dN}= & -3d_{1}+x-2(w-2)+y(5-2w)+y(w-2)-yx-\eta \\
       \frac{dz_{1}}{dN}= & -3bz_{1}-3z_{1}-2z_{1}(w-2)-z_{1}x\\
       \frac{dz_{2}}{dN}= & -3bz_{2}-4z_{2}-2z_{2}(w-2)-z_{2}x
   \end{split}
 \end{equation}
 
The above system represents the general dynamical system of $f(R,G,T)$ specified by $\frac{\ddot{f_{R}}}{H^{2}f_{R}}$. For our present model, $a$ and $b$ are given by
\begin{equation}\label{ab}
  a=l-1, \quad b=n-1.
\end{equation}
Also,
\begin{equation}\label{v}
  v=\frac{w}{l}+\frac{u}{m}-\frac{z_{1}}{2n}
\end{equation}
\begin{equation}\label{y}
  y=\frac{(m-1)u}{(w-1)^{2}}\big(\frac{wx}{l-1}+2(w-2)^{2}\big)
\end{equation}
\begin{equation}\label{x}
  x=\frac{-1+d_{1}+d_{2}+\frac{2n-1}{2n}z_{1}+\frac{4}{3}z_{2}+\frac{l-1}{l}w+\frac{(m-1)u}{(w-1)^{2}}\big(\frac{(w-1)^{2}}{m}-2(w-2)^{2}\big)}{1+\frac{(m-1)wu}{(l-1)(w-1)^{2}}}
\end{equation}

As the variables $v$, $y$, and $x$ are depending on others variables, the dynamical system (\ref{DS}) reduces to 6 (six) equations only.
\begin{equation}\label{DSF}
   \begin{split}
      \frac{dd_{1}}{dN}= & -3d_{1}-2d_{1}(w-2)-d_{1}x \\
      \frac{dd_{2}}{dN}= & -4d_{2}-2d_{2}(w-2)-d_{2}x \\
      \frac{du}{dN}=  & y(w-1)+\frac{u}{w-1}\big(\frac{wx}{l-1}+2(w-2)^{2}\big)-2u(w-2)-ux \\
       \frac{dw}{dN}= & \frac{wx}{l-1}-2w(w-2) \\
       \frac{dz_{1}}{dN}= & -3(n-1)z_{1}-3z_{1}-2z_{1}(w-2)-z_{1}x\\
       \frac{dz_{2}}{dN}= & -3(n-1)z_{2}-4z_{2}-2z_{2}(w-2)-z_{2}x
   \end{split}
 \end{equation}
 with $l\neq1,0$, $m\neq0$ and $n\neq0$.

Using (\ref{DEeos}) and (\ref{eff}), the equation of state for dark energy and effective equation of state are given by
\begin{equation}\label{DEeos1}
  \omega_{de}=-1-\frac{3d_{1}+4d_{2}+2w-4}{3(1-d_{1}-d_{2})}, \quad \omega_{eff}=-\frac{1}{3}(2w-1).
\end{equation}
Again, using (\ref{density}), density parameters for the matter, radiation and dark energy are defined by 
\begin{equation}\label{density1}
  \Omega_{m}=d_{1}, \quad \Omega_{r}=d_{2}, \quad \Omega_{de}=1-d_{1}-d_{2}.
\end{equation}
Lastly, the deceleration parameter, $q$ is given by
\begin{equation}\label{q}
  q=-1-\frac{\dot{H}}{H^{2}}=1-w.
\end{equation}

In the next section, we find the critical points of the dynamical system (\ref{DSF}) and analyse their stability. 

\section{Critical points of the dynamical system and their stability}\label{sec4}
 For finding the critical points of the dynamical system, all the equations involved in the dynamical system (\ref{DSF}) must be equate to zero. The critical points of the above dynamical system are given in the Tables (\ref{critical points}).

\begin{table}[h!]
\centering
\caption{Critical points (CP).}
\label{critical points}
    \centering
    \begin{tabular}{|l|l|l|l|l|l|l|l|}
        \hline
       CP & $d_{1}$ & $d_{2}$ & $u$ & $w$ & $z_{1}$ & $z_{2}$ & Existence\\
       \hline
       A & 0 & 1 & 0 & 0 & 0 & 0 & For all \\
        \hline
      B & 2 & 0 & 0 & 0 & 0 & 0 & For all \\
        \hline
      C & 0 & 0 & 0 & $\frac{4l-3n}{2l}$ & $C_{1}$ & 0 & $l\neq0$,$n\neq\frac{1}{2}$  \\
        \hline
      D & 0 & 0 & $\frac{2m-lm}{l(m-1)}$ & 2 & 0 & 0 & $l\neq0$,$m\neq1$  \\
         \hline
      E & 0 & $\frac{-(5l^2-8l+2)}{l^{2}}$ & 0 & $\frac{2(l-1)}{l}$ & 0 & 0 & $l\neq0$  \\
        \hline
      F & 0 & 0 & 0 & $\frac{l(4l-5)}{(2l-1)(l-1)}$ & 0 & 0 & $l\neq\frac{1}{2},1$  \\
        \hline
      G & 0 & 0 & 0 & $\frac{4l-3n-1}{2l}$ & 0 & $G_{1}$ & $l\neq0$  \\
      \hline
      H & $\frac{-8l^{2}+13l-3}{2l^{2}}$ & 0 & 0 & $\frac{4l-3}{2l}$ & 0 & 0 & For all \\
        \hline
    \end{tabular}
\end{table}

where $G_{1}=\frac{3(-1+7l-4l^{2}-3n+9ln-6l^{2}n)}{8l^{2}}$ and $C_{1}=\frac{n(4l-2l^{2}-3n+9ln-6l^{2}n)}{l^{2}(2n-1)}$.

For the dynamical system (\ref{DSF}), we get $eight$ critical points. The stability of these critical points can be analysed using the linear stability theory. The stability of these critical points and their cosmological interpretation are briefly explained below:

\begin{center}
Point A $(0,1,0,0,0,0)$
\end{center}

\begin{wrapfigure}{r}{0.4\textwidth}
  \centering
  \includegraphics[width=0.3\textwidth]{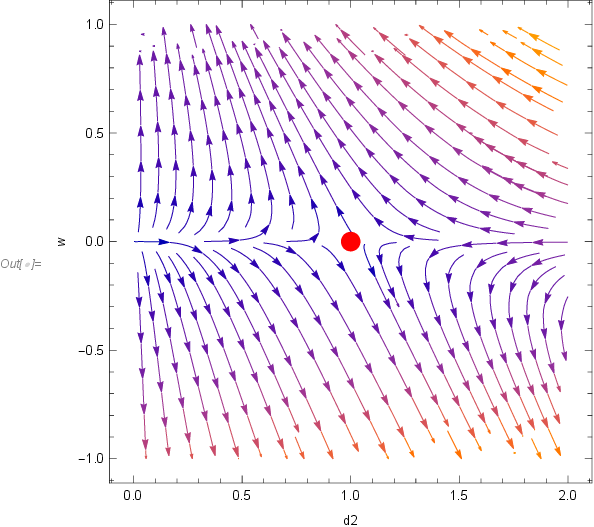}
   \caption{Phase portrait with $l=2$}
    \label{FA}
\end{wrapfigure}
This critical point exists for any value of the model's parameters. The eigenvalues of this point are $-1$, $1$, $4$ $-4(2m-1)$, $-3(n-1)$, and $4-3n$. As there exists at least one positive and negative eigenvalues, the present critical point is an unstable saddle point. Cosmic parameters such as density parameters, equation of state and deceleration parameter are given by
\begin{equation}\label{paraA}
  \Omega_{m}=0, \quad \Omega_{r}=1, \quad \Omega_{de}=0, \quad \omega_{de}=\infty,\quad \omega_{eff}=\frac{1}{3}, \quad q=1.
\end{equation} \\

The Hubble parameter and the scale factor are given by
\begin{equation}\label{A}
  H=\frac{1}{2t+C}, \quad a=K\sqrt{2t+C},
\end{equation}
where $C$ and $K$ are the constants of integration. From (\ref{paraA}), it is clear that this point represents the phase of the deceleration expansion and further suggests that the Universe is dominated by radiation. The corresponding phase portrait when $l=2$, is given by Figure \ref{FA}.

\begin{center}
Point B $(2,0,0,0,0,0)$ 
\end{center}
This critical point represent the matter dominated era of the Universe as the values of $\Omega_{m}=2$, $\Omega_{r}=0$ and $\Omega_{de}=-1$. The Hubble parameter and the scale factor are given by 
\begin{equation}\label{B}
  H=\frac{1}{2t+C}, \quad a=K\sqrt{2t+C},
\end{equation}
where $C$ and $K$ are the constants of integration. By equation (\ref{B}), it is clear that $a(t)\propto t^{\frac{1}{2}}$, which is a contraction. As, for a matter-dominated Universe $a(t)\propto t^{\frac{2}{3}}$, which is obtained by solving the Friedmann equations. Hence, this critical point can be ruled out as it is non physical.

\begin{center}
Point C $(0,0,0,\frac{4l-3n}{2l},\frac{n(4l-2l^{2}-3n+9ln-6l^{2}n)}{l^{2}(2n-1)},0)$
\end{center}

The corresponding eigenvalues of this CP are $-1$, $3(n-1)$, $\frac{3n(2m-l)}{l}$, $3n-4$, \\
$\frac{3(2ln-n-2l)}{4l}-\frac{1}{4}\sqrt{\frac{-164 l^2 + 100 l^3 + 156 l n - 228 l^2 n + 120 l^3 n - 81 n^2 + 261 l n^2 - 216 l^2 n^2 + 36 l^3 n^2}{(-1 + l) l^2}}$ and\\ $\frac{3(2ln-n-2l)}{4l}+\frac{1}{4}\sqrt{\frac{-164 l^2 + 100 l^3 + 156 l n - 228 l^2 n + 120 l^3 n - 81 n^2 + 
 261 l n^2 - 216 l^2 n^2 + 36 l^3 n^2}{(-1 + l) l^2}}$. As one of the eigenvalues is negative, this CP cannot be unstable. This CP is stable for $l>0$, $0<n<1$, $m>1$ with the additional inequality $(4 l - 3 n) (-4 l + 2 l^2 + 3 n - 9 l n + 6 l^2 n)<0$ or $\frac{-164 l^2 + 100 l^3 + 156 l n - 228 l^2 n + 120 l^3 n - 81 n^2 + 
 261 l n^2 - 216 l^2 n^2 + 36 l^3 n^2}{(-1 + l) l^2}<0$. Otherwise this CP is a saddle point. Cosmic parameters such as density parameters, equation of state and deceleration parameter are given by
\begin{equation}\label{paraC}
\begin{split}
  &\Omega_{m}=0, \quad \Omega_{r}=0, \quad \Omega_{de}=1, \quad \omega_{de}=\frac{-(l-n)}{l},\\
  &\quad \omega_{eff}=\frac{-(l-n)}{l}, \quad q=\frac{-(2l-3n)}{2l}.
  \end{split}
\end{equation} \\

\begin{wrapfigure}{r}{0.4\textwidth}
  \centering
  \includegraphics[width=0.3\textwidth]{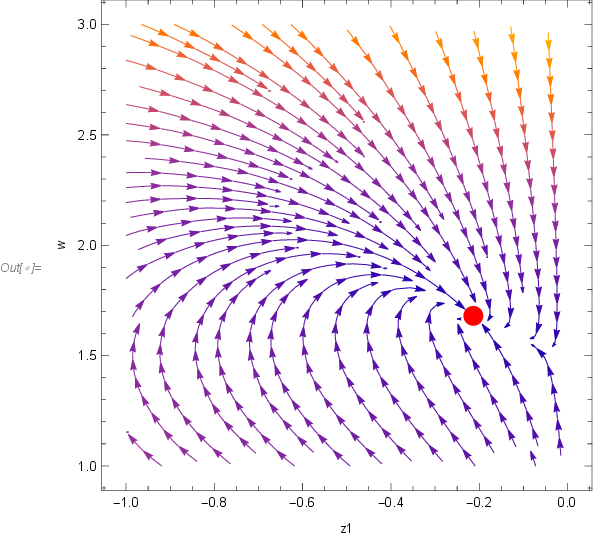}
   \caption{Phase portrait with $l=2$ and $n=1$}
    \label{FC}
\end{wrapfigure}

The Hubble parameter and the scale factor are given by
\begin{equation}\label{C}
  H=\frac{1}{\frac{3nt}{2l}+C}, \quad a=K\bigg(\frac{3nt}{2l}+C\bigg)^{\frac{2l}{3n}},
\end{equation}
where $C$ and $K$ are the constants of integration. From (\ref{paraC}), it is clear that this point represents the phase of the accelerated expansion and further suggests that the Universe is dominated by dark energy for $2l>3n$. For $l=1.5$ and $n=0.32$, $\omega_{eff}=-0.787$ and $q=-0.68$, which indicate that the Universe is in the phase of Quintessence and the expansion of the Universe is accelerating. Figure \ref{FC} shows the corresponding phase portrait for $l=1.5$ and $n=0.32$ and conclude that at this values of the parameters this CP is an unstable point.  

\begin{center}
Point D $(0,0,\frac{2m-lm}{l(m-1)},2,0,0)$
\end{center}

  The eigenvalues of this point are $-4$, $-3$, $-3 n$, $-3 n-1$, $\frac{-3}{2} - \frac{1}{2}\sqrt{25-\frac{16l}{-l + l^2 + 4 m - 2 l m}}$ and $\frac{-3}{2} + \frac{1}{2}\sqrt{25-\frac{16l}{-l + l^2 + 4 m - 2 l m}}$. A critical point is stable if all the corresponding eigenvalues are negative. So, this critical point is stable if $-41l+25l^{2}+100m-50lm<0$ and $n>0$. Also for $0<\sqrt{25-\frac{16l}{-l + l^2 + 4 m - 2 l m}}<9$, the CP is stable. Otherwise this CP is a saddle point.  Cosmic parameters such as density parameters, equation of state and deceleration parameter are given by
\begin{equation}\label{paraD}
  \Omega_{m}=0, \quad \Omega_{r}=0, \quad \Omega_{de}=1, \quad \omega_{de}=-1,\quad \omega_{eff}=-1, \quad q=-1.
\end{equation} \\ 
\begin{wrapfigure}{r}{0.4\textwidth}
  \centering
  \includegraphics[width=0.3\textwidth]{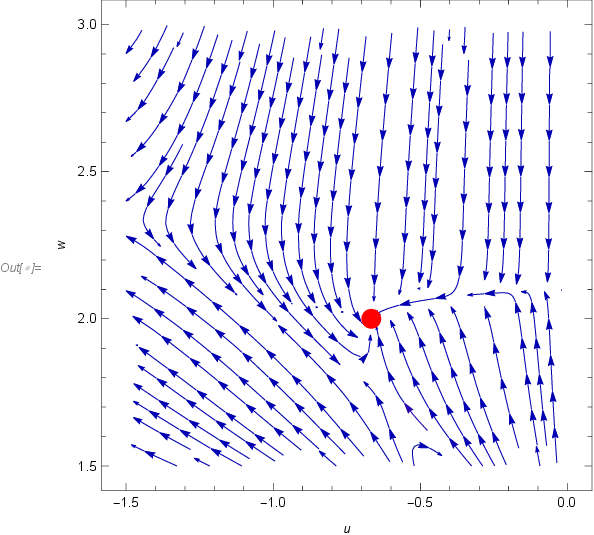}
   \caption{Phase portrait with $l=3$, $m=2$ and $n=2$}
    \label{FD}
\end{wrapfigure}
\\
The Hubble parameter and the scale factor are given by
\begin{equation}\label{C}
  H=C, \quad a=Ke^{Ct},
\end{equation}
where $C$ and $K$ are the constants of integration. From (\ref{paraD}), it is clear that this point represents the phase of the accelerated expansion and further suggests that the Universe is dominated by dark energy. This CP represents a de-Sitter model of the Universe. The corresponding phase portrait when $l=3$, $m=2$ and $n=2$, is given by Figure \ref{FD}.

\begin{center}
Point E $(0,\frac{-(5l^2-8l+2)}{l^{2}},0,\frac{2(l-1)}{l},0,0)$
\end{center}

  The eigenvalues of this point are $1$, $\frac{4(l-2m)}{l}$, $-3(n-1)$, $4-3 n$, $\frac{l-2}{2l} - \frac{\sqrt{3(12 - 44 l + 27 l^2)}}{ 2l}$ and $\frac{l-2}{2l} + \frac{\sqrt{3(12 - 44 l + 27 l^2)}}{ 2l}$. As one of the eigenvalue is positive, this CP will be either a saddle point or an unstable point, (if all eigenvalues have positive real part). So, this critical point is saddle if $0<l<2m$ or $n>1$ . Also, this CP can not be unstable as the last two eigenvalues cannot have positive real part at the same time. 
  
  \begin{wrapfigure}{r}{0.4\textwidth}
  \centering
  \includegraphics[width=0.3\textwidth]{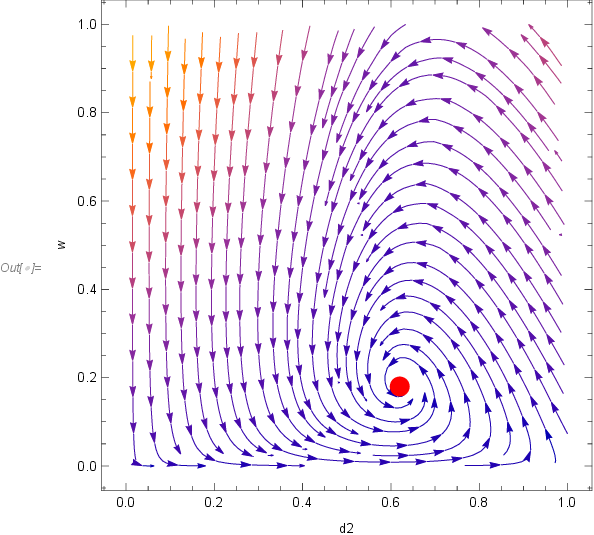}
   \caption{Phase portrait with $l=1.1$}
    \label{FE}
\end{wrapfigure}

 Cosmic parameters such as density parameters, equation of state and deceleration parameter are given by
\begin{equation}\label{paraE}
\begin{split}
  &\Omega_{m}=0, \quad \Omega_{r}=\frac{-2+8l-5l^2}{l^2}, \quad \Omega_{de}=\frac{2-8l+6l^2}{l^2},\\
  & \quad \omega_{de}=\frac{\frac{20}{3}l-l^2}{2-8l+6l^2},\quad \omega_{eff}=\frac{-(3l-4)}{3l}, \quad q=\frac{-(l-2)}{l}.
  \end{split}
\end{equation} \\ 
The Hubble parameter and the scale factor are given by
\begin{equation}\label{E}
  H=\frac{1}{\frac{2t}{l}+C}, \quad a=K\big(\frac{2t}{l}+C\big)^{\frac{l}{2}},
\end{equation}
where $C$ and $K$ are the constants of integration. From (\ref{paraE}), it is clear that both $\Omega_{r}$ and $\Omega_{de}$ are positive for $1<l<1.2899$. This point represents the phase of the Universe dominated by both radiation and dark energy. The corresponding phase portrait when $l=1.1$, is given by Figure \ref{FE}. At $l=1.1$, the last two eigenvalues occur as a pair of complex conjugate with negative real part, making the CP spiral in saddle point.

\begin{center}
Point F $(0,0,0,\frac{l(4l-5)}{(2l-1)(l-1)},0,0)$
\end{center}

  The eigenvalues of this point are $\frac{-(4l-5)}{l-1}$, $\frac{-2 (2 - 8 l + 5 l^2)}{(2l-1)(l-1)}$, $\frac{-(3 - 13 l + 8 l^2)}{(2l-1)(l-1)}$, $\frac{-2 (-2 + l) (l - 2 m)}{(2l-1)(l-1)}$, $\frac{-(-4 l + 2 l^2 + 3 n - 9 l n + 6 l^2 n)}{(2l-1)(l-1)}$, and  $\frac{-(1-7 l + 4 l^2 + 3 n - 9 l n + 6 l^2 n)}{(2l-1)(l-1)}$. This CP is stable if $l\notin[0.3,1.3]$, $(l-2m)(l-2)>0$, $n>max\{\frac{-2l(l-2)}{3(2l-1)(l-1)},\frac{-(4l^2-7l+1)}{3(2l-1)(l-1)}\}$.
  \\
 Cosmic parameters such as density parameters, equation of state and deceleration parameter are given by
\begin{equation}\label{paraF}
\begin{split}
  &\Omega_{m}=0, \quad \Omega_{r}=0, \quad \Omega_{de}=1, \quad \omega_{de}=\frac{-(6l^2-7l-1)}{3(2l-1)(l-1)}, \quad \omega_{eff}=\frac{-(6l^2-7l-1)}{3(2l-1)(l-1)}, \quad q=\frac{-(2l^2-2l-1)}{(2l-1)(l-1)}.
  \end{split}
\end{equation} \\ 

 \begin{wrapfigure}{r}{0.4\textwidth}
  \centering
  \includegraphics[width=0.3\textwidth]{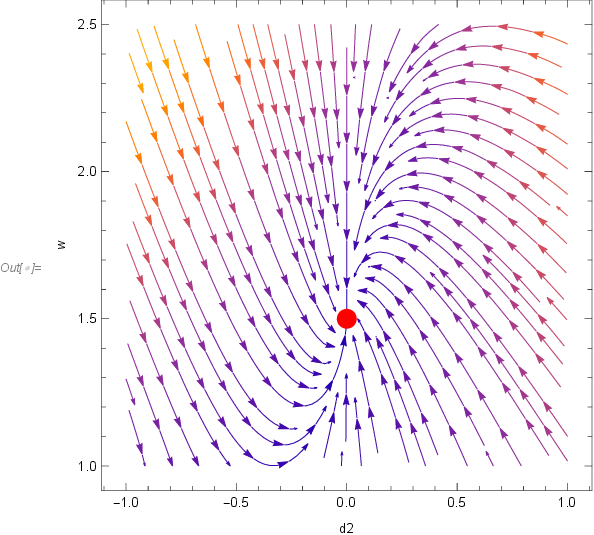}
   \caption{Phase portrait with $l=1.5$}
    \label{FF}
\end{wrapfigure}

The Hubble parameter and the scale factor are given by
\begin{equation}\label{F}
  H=\frac{1}{\frac{(l-2)t}{(2l-1)(l-1)}+C}, \quad a=K\bigg(\frac{(l-2)t}{(2l-1)(l-1)}+C\bigg)^{\frac{(2l-1)(l-1)}{(l-2)}},
\end{equation}
where $C$ and $K$ are the constants of integration. From (\ref{paraF}), it is clear that the Universe is dominated by dark energy. At $l=1.5$, $q=-0.5$ and $\omega_{eff}=-\frac{2}{3}$, which corresponds to Quintessence stage of the Universe and the Universe's expansion is accelerating. Figure \ref{FF} represents the corresponding phase portrait when $l=1.5$.

\begin{center}
Point G $(0,0,0,\frac{4l-3n-1}{2l},0,\frac{3(-1+7l-4l^{2}-3n+9ln-6l^{2}n)}{8l^{2}})$
\end{center}

The corresponding eigenvalues of this CP are $1$, $3(n-1)$, $\frac{(2m-l)(3n+1)}{l}$, $3n-2$,\\ $\frac{-1 - 4 l - 3 n + 6 l n}{4 l}-\frac{\sqrt{3}}{4}\sqrt{\frac{-3 + 27 l - 88 l^2 + 48 l^3 - 18 n + 110 l n - 124 l^2 n + 48 l^3 n -  27 n^2 + 87 l n^2 - 72 l^2 n^2 + 12 l^3 n^2}{(-1 + l) l^2}}$ and \\ $\frac{-1-4l-3n+6ln}{4l}+\frac{\sqrt{3}}{4}\sqrt{\frac{-3 + 27 l - 88 l^2 + 48 l^3 - 18 n + 110 l n - 124 l^2 n + 48 l^3 n - 27 n^2 + 87 l n^2 - 72 l^2 n^2 + 12 l^3 n^2}{(-1 + l) l^2}}$. 
\begin{wrapfigure}{r}{0.4\textwidth}
  \centering
  \includegraphics[width=0.3\textwidth]{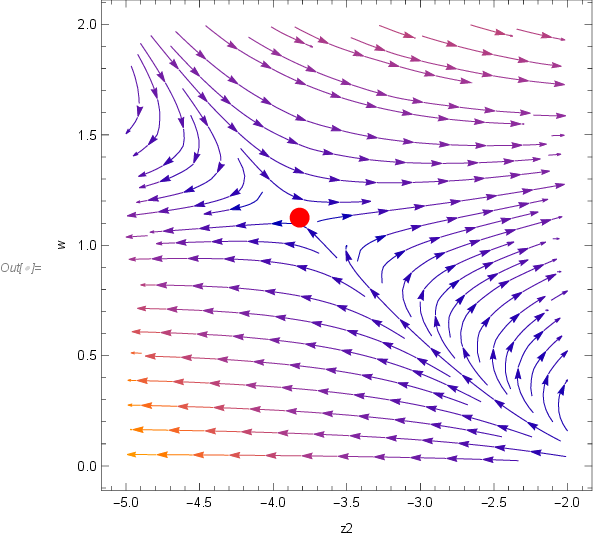}
   \caption{Phase portrait with $l=2$ and $n=1.5$}
    \label{FG}
\end{wrapfigure}

Here one of the eigenvalues is positive, so this CP cannot be a stable point. This CP is unstable if $n>1$, $2m>l$ and $l>\frac{3n+1}{6n-4}$ satisfying \\ $\frac{-3 + 27 l - 88 l^2 + 48 l^3 - 18 n + 110 l n - 124 l^2 n + 48 l^3 n - 
 27 n^2 + 87 l n^2 - 72 l^2 n^2 + 12 l^3 n^2}{l-1}<0$ or $n>1$, $2m>l$, $l>\frac{3n+1}{6n-4}$ and $\frac{(1 - 4 l + 3 n) (1 - 7 l + 4 l^2 + 3 n - 9 l n + 6 l^2 n)}{l-1}>0$, otherwise it is a saddle point. Cosmic parameters such as density parameters, equation of state and deceleration parameter are given by
\begin{equation}\label{paraG}
\begin{split}
  &\Omega_{m}=0, \quad \Omega_{r}=0, \quad \Omega_{de}=1, \quad \omega_{de}=\frac{-(3l-3n-1)}{3l},\\
  &\quad \omega_{eff}=\frac{-(3l-3n-1)}{3l}, \quad q=\frac{-(2l-3n-1)}{2l}.
  \end{split}
\end{equation} \\

The Hubble parameter and the scale factor are given by
\begin{equation}\label{G}
  H=\frac{1}{\frac{(3n+1)t}{2l}+C}, \quad a=K\bigg(\frac{(3n+1)t}{2l}+C\bigg)^{\frac{2l}{3n+1}},
\end{equation}
\begin{wrapfigure}{r}{0.4\textwidth}
  \centering
  \includegraphics[width=0.3\textwidth]{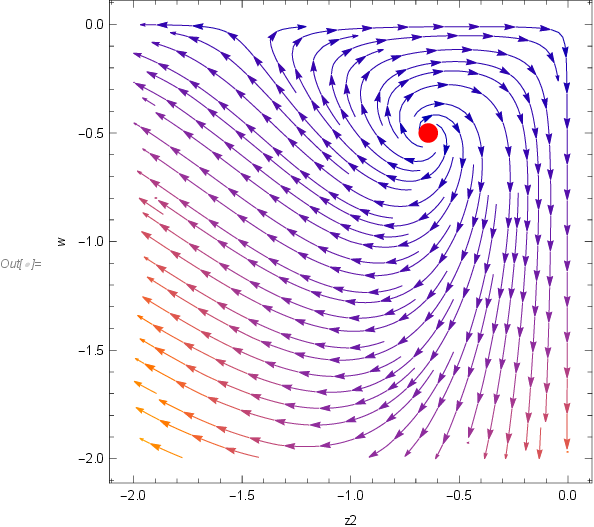}
   \caption{Phase portrait with $l=2$ and $n=1.5$}
    \label{FGu}
\end{wrapfigure}
where $C$ and $K$ are the constants of integration. From (\ref{paraG}), it is clear that this point represents the phase of the accelerated expansion and further suggests that the Universe is dominated by dark energy for $2l>3n+1$. For $l=4$ and $n=2$, $\omega_{eff}=\frac{-5}{12}$ and $q=\frac{-1}{8}$, which indicate that the Universe is in the phase of Quintessence and the Universe's expansion is accelerating. Figure \ref{FG} shows the corresponding phase portrait for $l=4$ and $n=2$ for an unstable saddle point. Figure \ref{FGu} shows the corresponding phase portrait for $l=0.5$ and $n=1.1$ for spiral unstable point.

\begin{center}
Point H $(\frac{-8l^{2}+13l-3}{2l^{2}},0,0,\frac{4l-3}{2l},0,0)$
\end{center}

\begin{wrapfigure}{l}{0.4\textwidth}
  \centering
  \includegraphics[width=0.3\textwidth]{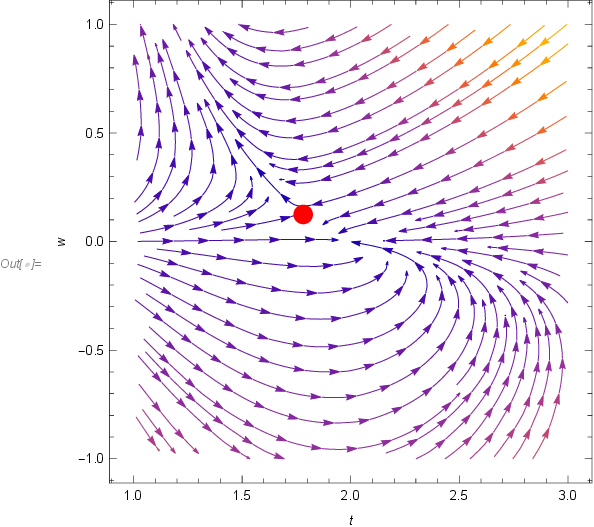}
   \caption{Phase portrait with $l=0.8$}
    \label{FHs}
\end{wrapfigure}

\begin{wrapfigure}{r}{0.4\textwidth}
  \centering
  \includegraphics[width=0.3\textwidth]{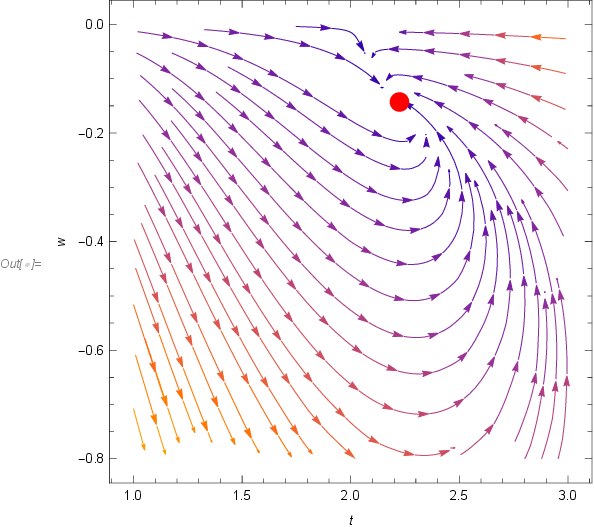}
   \caption{Phase portrait with $l=0.7$}
    \label{FHss}
\end{wrapfigure}

This critical point exists for any value of the model's parameters. The eigenvalues of this point are $-1$, $\frac{3(2m-l)}{l}$, $2-3n$, $-3(n-1)$, $-\frac{3}{4l}-\frac{\sqrt{(-1 + l) (-81 + 417 l - 608 l^2 + 256 l^3)}}{4l(l-1)}$, and $-\frac{3}{4l}+\frac{\sqrt{(-1 + l) (-81 + 417 l - 608 l^2 + 256 l^3)}}{4l(l-1)}$. As there exists at least negative eigenvalue, the present critical point is a stable or a saddle point. The condition for the sable is that $l\in (0.2785,0.75) \bigcup (1,1.3465)$, $2m>l$ and $n>1$. Otherwise, this CP is a saddle point. Cosmic parameters such as density parameters, equation of state and deceleration parameter are given by
\begin{equation}\label{paraH}
\begin{split}
 & \Omega_{m}=\frac{-8l^{2}+13l-3}{2l^{2}}, \quad \Omega_{r}=0, \quad \Omega_{de}=\frac{10l^{2}-13l+3}{2l^{2}}, \\
 & \omega_{de}=\frac{-l(2l-3)}{(10l-3)(l-1)},\quad \omega_{eff}=\frac{-(l-1)}{l}, \quad q=\frac{-(2l-3)}{2l}.
  \end{split}
\end{equation} 

The Hubble parameter and the scale factor are given by
\begin{equation}\label{H}
  H=\frac{1}{\frac{3t}{2l}+C}, \quad a=K\bigg(\frac{3t}{2l}+C\bigg)^\frac{2l}{3},
\end{equation}
where $C$ and $K$ are the constants of integration. From (\ref{paraH}), it is clear that this point represents the phase of the Universe dominated by matter or dark energy or both. For $l>1.5$, the Universe is in the phase of accelerated expansion while for $l<1.5$, it is in the phase of decelerated expansion. For $l=1.5$, the Universe is in the transition phase. The corresponding phase portrait when $l=0.8$ and $l=0.7$, are given by Figure \ref{FHs} and \ref{FHss}.

\begin{figure}[h!]
  \centering
  \includegraphics[width=0.6\textwidth]{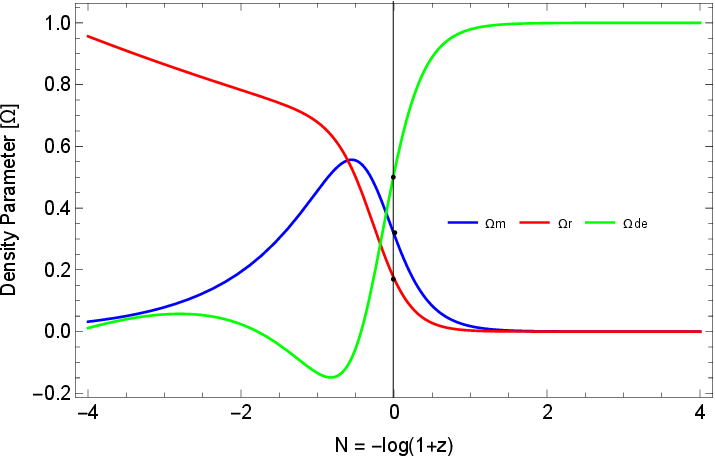}
  \caption{Evolution of density parameters with initial condition $d1=0.315$, $d2=0.171$ and $w=1.50$}\label{density-parameters}
\end{figure}

The evolution of density parameters when $x=0.1$ and $l=2$ is shown in figure (\ref{density-parameters}). It also show that the evolution is consistent with the current observation with present value calculated as $\Omega_{m}\approx 0.315$, $\Omega_{r}\approx 0.171$ and $\Omega_{de}\approx 0.514$. Initially, there is domination of radiation in the early phase of Universe, followed by matter dominance and in the late time, dark energy is dominating the Universe (de-Silter), leading to the late time acceleration.\\

\section{Cosmic parameters}\label{cosmic-p}

\begin{figure}[htbp]
  \centering
  \begin{subfigure}[b]{0.32\textwidth}
    \includegraphics[width=1.2\linewidth]{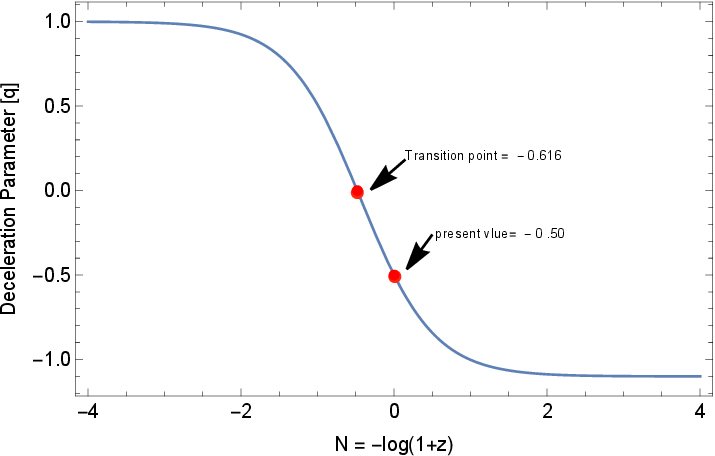}
    \caption{Plot of deceleration with respect to $N=-log(1+z)$}
    \label{fig:q}
  \end{subfigure}
  \hfill
  \begin{subfigure}[b]{0.32\textwidth}
    \includegraphics[width=1.2\linewidth]{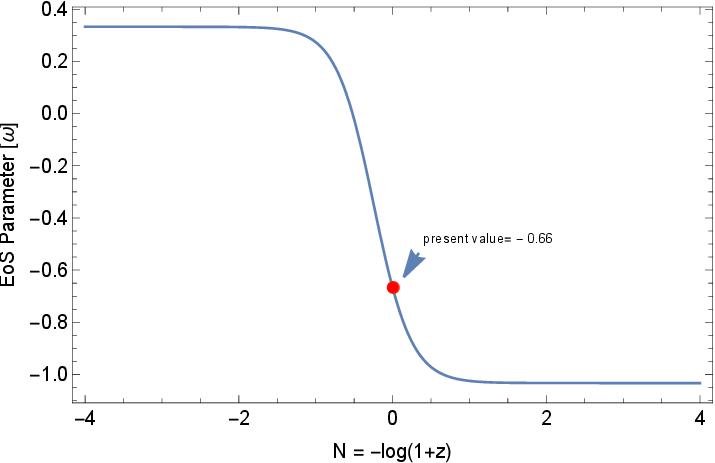}
    \caption{Plot of $\omega$ with respect to $N=-log(1+z)$}
    \label{fig:oemga}
  \end{subfigure}
  \caption{Plot of cosmic parameters with initial condition $w=1.50$ vs $N=-log(1+z)$ }
\end{figure}

\textbf{Deceleration parameter:} It measures the rate of change of the expansion of the universe. A negative value, positive value and zero value of $q$ mean that the Universe's expansion is accelerating, decelerating and constant respectively. The evolution of $q[N]$ with respect to $N=-log(1+z)$ for $l=2$ is given in figure (\ref{fig:q}). The deceleration parameter starts from positive value, deceleration phase and goes to negative value, acceleration phase. At $N=-0.48$ i.e $z_{tr}=e^{-N}-1=0.616$, there is a transition from deceleration phase to acceleration phase of the Cosmos, which is align with the observational constraints $0.5\leq z_{tr}\leq0.8$. The present value of $q=-0.50$ indicates that currently our Universe is in the phase of accelerated expansion.

\textbf{Equation of state:} It describes the relationship between the pressure and energy density of different component of the Universe, $\omega=\frac{p}{\rho}$. The evolution of $\omega[N]$ with respect to $N=-log(1+z)$ for $l=2$ is given in figure (\ref{fig:oemga}). The equation of state parameter starts with a positive value, $\omega \approx 0.3$ (dominated by matter in the early universe), and goes to negative value (dominated by dark energy in the present and late-time Universe). The present value of $\omega \approx -0.66$, indicates that our Universe is in the phase of Quintessence.

\textbf{The state finder parameters:} The state finder parameters $\{r,s\}$ act as sophisticated diagnostic tools designed to distinguish between various dark energy (DE) models by examining their dynamic behavior \cite{V. Sahni}. In contrast to standard cosmological indicators like the Hubble and deceleration parameters—which rely mainly on lower-order derivatives of the scale factor $a(t)$—the state finder approach incorporates higher-order derivatives, offering a more detailed perspective on the Universe's expansion history.

The state finder parameter $r$, the jerk parameter, is given by
\begin{equation}
    r = q(1+2q)-\frac{dq}{dN}=\frac{wx}{l-1}-w+3
\end{equation}

 This parameter $r$ characterizes the rate of change of cosmic acceleration and serves as a crucial discriminator between dark energy models.

Additionally, the parameter $s$, the snap parameter, is given as
\begin{equation}
    s = \frac{r - 1}{3(q - 1/2)}=\frac{wx+(l-1)(2-w)}{3(l-1)(\frac{1}{2}-w)}
\end{equation}
 As $s$ is related on both $q$ and $r$, it provides insights into the evolution of cosmic acceleration and also the nature of DE.
The state finder approach allows for the categorization of cosmological models independently of their underlying theoretical assumptions. The different values $\{r=1,s=0\}$, $\{r<1,s>0\}$ and $\{r>1,s<0\}$ denote $\Lambda$CDM, Quintessence and Chaplygin Gas model of dark energy respectively.

\begin{table}[h!]
\centering
\begin{tabular}{|c|c|c|c|c|c|}
\hline
CP & $q$ & $\omega_{eff}$ & $r$ & $s$ & Model  \\
\hline
A   & $1$   & $\frac{1}{3}$   &  $3$  & $\frac{4}{3}$ & Radiation dominated phase  \\ 
\hline
C & $\frac{-(2l-3n)}{2l}$ & $\frac{-(l-n)}{l}$ & $\frac{(l-3n)(2l-3n)}{2l^2}$ & $\frac{n}{l}$ &  Quintessence, $l>n>0$ \\
\hline
D & $-1$ & $-1$  & $1$ & $0$ & $\Lambda$CDM \\
\hline
E & $\frac{-(l-2)}{l}$ & $\frac{-(3l-4)}{3l}$ & $\frac{(l+1)(l^2-2l+2)}{l^3}$ & $\frac{2(l^2-2)}{3l^2(3l-4)}$ & Quintessence,$l>\sqrt{2}$ and $\Lambda$CDM,$l=\sqrt{2}$   \\
\hline
F & $\frac{-(2l^{2}-2l-1)}{(2l-1)(l-1)}$ & $\frac{-(6l^{2}-7l-1)}{3(2l-1)(l-1)}$ & $\frac{(l+1)(2l-3)(2l^{2}-2l-1)}{(2l-1)^{2}(l-1)^{2}}$ & $\frac{-2(l-2)}{3(2l-1)(l-1)}$ & $\Lambda$CDM,$l=2$; Chaplygin gas,$l>2$ \\
\hline
G  & $\frac{-(2l-3n-1)}{2l}$ & $\frac{-(3l-3n-1)}{2l}$ & $\frac{(l-3n-1)(2l-3n-1)}{2l^2}$ & $\frac{3n+1}{3l}$ &  Quintessence, $l-n>\frac{1}{3}$  \\
\hline
H & $\frac{-(2l-3)}{2l}$ & $\frac{-(l-1)}{l}$ &  $\frac{(l-3)(2l-3)}{2l^2}$ & $\frac{1}{l}$ & Quintessence, $l>1$  \\
\hline
\end{tabular}
\caption{Cosmic parameters for the model $f(R,G,T)=\alpha R^{l}+\beta G^{m} +\gamma T^{n}$}
\label{tab:parameters}
\end{table}

\section{Validity of the proposed model using observational datasets}\label{comparison}

\begin{figure}[htbp]
  \centering
  \begin{subfigure}[b]{0.32\textwidth}
    \includegraphics[width=\linewidth]{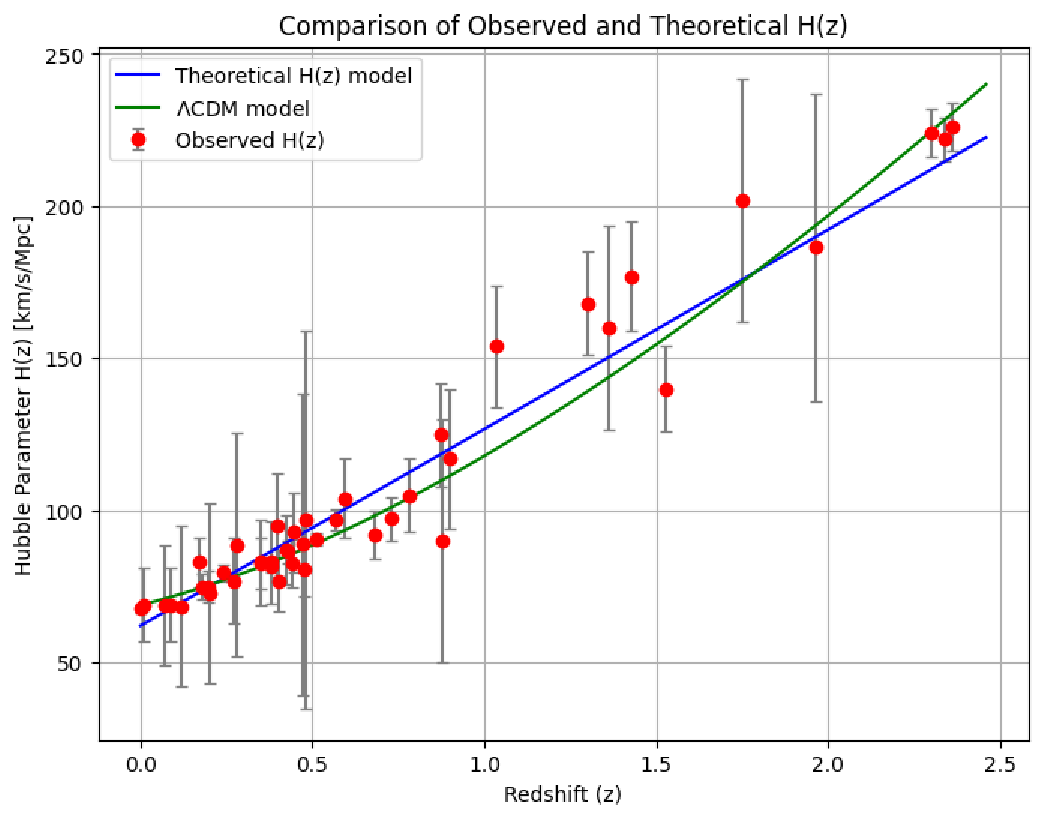}
    \caption{comparison with Hubble Data}
    \label{fig:hubble}
  \end{subfigure}
  \hfill
  \begin{subfigure}[b]{0.32\textwidth}
    \includegraphics[width=\linewidth]{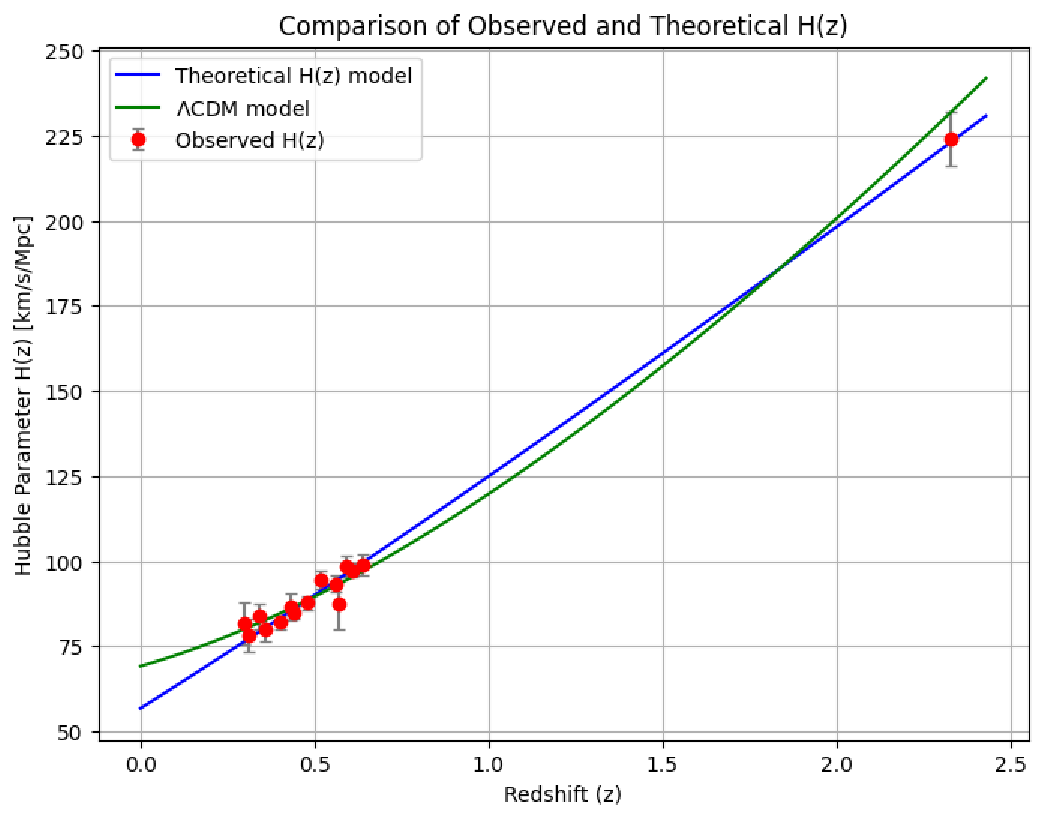}
    \caption{Comparison with BAO Data}
    \label{fig:bao}
  \end{subfigure}
  \hfill
  \begin{subfigure}[b]{0.32\textwidth}
    \includegraphics[width=1.25\linewidth]{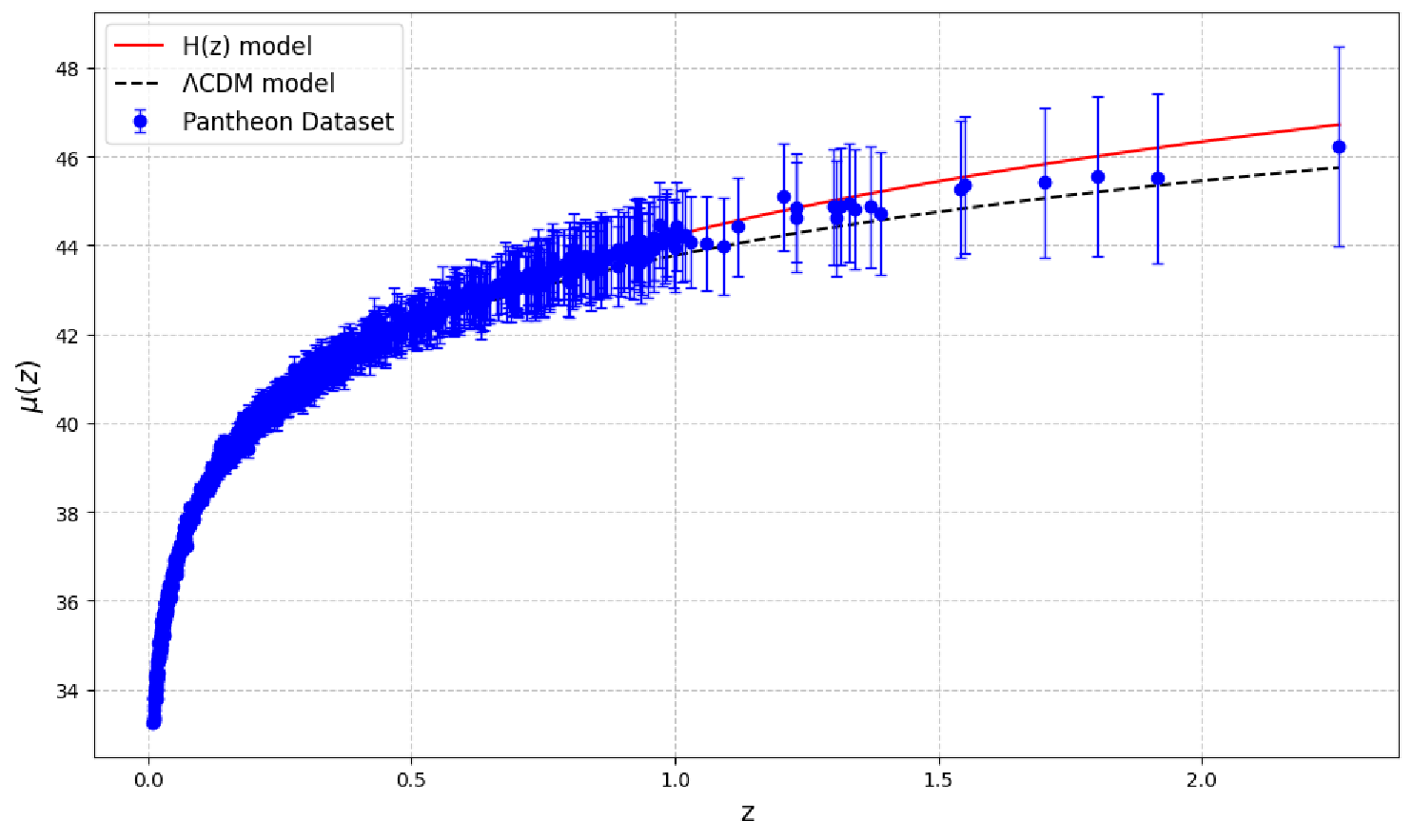}
    \caption{Comparison with pantheon data}
    \label{fig:pantheon}
  \end{subfigure}
  \caption{Comparison with observational datasets}
\end{figure}

In this section, we check the validity of the proposed model $f(R,G,T)=\alpha R^{l}+\beta G^{m}+\gamma T^{n}$, by taking the dimensionless parameter $w$ as scalar in the equation, $\frac{\dot{H}}{H^{2}}=w-2$. Then, Hubble parameter is obtained as $H(z)=H_{0}(1+z)^{2-w}$, where $H_{0}$ is the present value of $H(z)$. Validity of the model is checked by comparing to the model, $H(z)^{2}=H_{0}^{2}[\Omega_{m}(1+z)^{3}+\Omega_{\Lambda}+(1-\Omega_{m}-\Omega_{\Lambda})^4]$, where $\Omega_{m}$ and $\Omega_{\Lambda}$ represent the density parameter of matter and dark energy. We take $H_{0}=67.4$ km/s/Mpc, $\Omega_{m}=0.3153$ and $\Omega_{\Lambda}=0.6947$ as on \cite{Planck2018}. Also, we take $w=1.50$. The evolution of our model's Hubble parameter is tested with 77 CC dataset and 15 BAO dataset as shown in figure (\ref{fig:hubble}) and (\ref{fig:bao}) respectively, which show our Hubble parameter is evolved as that of the $\Lambda$CDM model. The modulus distance function of our model with that of the $\Lambda$CDM model is given in figure (\ref{fig:pantheon}) using 1048 pantheon dataset, indicating same behaviour. The modulus distance function is defined by $\mu=m-M=5log_{10}(d)+25$ or $d=10^{\frac{\mu-25}{5}}$, where $m$ is the apparent magnitude , $M$ is the absolute magnitude, $\mu$ is the distance modulus and $d$ is the distance to the object in parsecs. It is used in Type la supernovae to measure cosmic expansion. The alignment of our model and the observational data gives us strong agreement to the standard cosmological evidence. This datasets are highlighted at \cite{J.K. Singh,Amit s.}

\section{Results and discussion}\label{conculsion}

In this study, we consider the modified $f(R,G,T)$ gravity to develop a dynamical system of ordinary differential equations (ODEs). For this, we define $9$ dimensionless variables as given in section (\ref{sec3}) and find their derivatives with respect to the logarithmic time $N=log(a(t))$. To make the dynamical system a set of autonomous ODEs we used a specific form of the model i.e. $f(R,G,T)=\alpha R^{l}+\beta G^{m}+\gamma T^{n}$. With respect to this model, we obtained $8$ CPs as given in Table (\ref{critical points}), out of which the second critical point, $B$ is neglected due to its non physical scenario.
\begin{itemize}
  \item A is a saddle point for all the parameters in the model. It represented the phase of early inflation phase of the Universe dominated by radiation with $\omega_{eff}=\frac{1}{3}$ and $q=1$. Here, scale factor, $a(t)\propto t^\frac{1}{2}$. 
  \item C is stable under the condition $l>0$, $0<n<1$, $m>1$ with the additional inequality $(4 l - 3 n) (-4 l + 2 l^2 + 3 n - 9 l n + 6 l^2 n)<0$ or $\frac{-164 l^2 + 100 l^3 + 156 l n - 228 l^2 n + 120 l^3 n - 81 n^2 +  261 l n^2 - 216 l^2 n^2 + 36 l^3 n^2}{(-1 + l) l^2}<0$, otherwise unstable. It represents the phase of Universe dominated by dark energy with $\omega_{eff}=\frac{-(l-n)}{l}, \quad q=\frac{-(2l-3n)}{2l}$. Here, scale factor, $a(t)\propto t^\frac{2l}{3n}$.

  \item D is stable under the condition $-41l+25l^{2}+100m-50lm<0$ and $n>0$ and $0<\sqrt{25-\frac{16l}{-l + l^2 + 4 m - 2 l m}}<9$. It represents the phase of the Universe dominated by dark energy with $\omega_{eff}=-1, \quad q=-1$. Here, scale factor, $a(t)\propto e^t$.
      
  \item E is saddle under the condition $0<l<2m$ or $n>1$. It represents the phase of the Universe with both radiation and dark energy with $\omega_{eff}=\frac{-(3l-4)}{3l}, \quad q=\frac{-(l-2)}{l}$. Here, scale factor, $a(t)\propto t^\frac{l}{2}$.
      
  \item F is stable under the condition $l\notin[0.3,1.3]$, $(l-2m)(l-2)>0$, $n>max\{\frac{-2l(l-2)}{3(2l-1)(l-1)},\frac{-(4l^2-7l+1)}{3(2l-1)(l-1)}\}$. It shows phase of the Universe dominated by dark energy with $\omega_{eff}=\frac{-(6l^2-7l-1)}{3(2l-1)(l-1)}, \quad q=\frac{-(2l^2-2l-1)}{(2l-1)(l-1)}$. Here, scale factor, $a(t)\propto t^\frac{(2l-1)(l-1)}{l-2}$.
      
  \item G is unstable under the condition $n>1$, $2m>l$ and $l>\frac{3n+1}{6n-4}$ satisfying \\ $\frac{-3 + 27 l - 88 l^2 + 48 l^3 - 18 n + 110 l n - 124 l^2 n + 48 l^3 n - 
 27 n^2 + 87 l n^2 - 72 l^2 n^2 + 12 l^3 n^2}{l-1}<0$ or $n>1$, $2m>l$, $l>\frac{3n+1}{6n-4}$ and $\frac{(1 - 4 l + 3 n) (1 - 7 l + 4 l^2 + 3 n - 9 l n + 6 l^2 n)}{l-1}>0$, otherwise it is a saddle point. It highlights the phase of Universe dominated by dark energy with $\omega_{eff}=\frac{-(3l-3n-1)}{3l}, \quad q=\frac{-(2l-3n-1)}{2l}$. Here, scale factor, $a(t)\propto t^\frac{2l}{3n+1}$.
 
  \item H is stable under the condition $l\in (0.2785,0.75) \bigcup (1,1.3465)$, $2m>l$ and $n>1$, otherwise saddle. It gives the phase of the Universe with both matter and dark energy with $\omega_{eff}=\frac{-(l-1)}{l}, \quad q=\frac{-(2l-3)}{2l}$. Here, scale factor, $a(t)\propto t^\frac{2l}{3}$.
\end{itemize}
Analysis of the evolution of density parameters when $x=0.1$ and $l=2$ is shown in figure (\ref{density-parameters}). It also shows that the evolution is consistent with the current observation as present values given as $\Omega_{m}\approx 0.315$, $\Omega_{r}\approx 0.171$ and $\Omega_{de}\approx 0.514$. Initially, there is domination of radiation in the early phase of Universe, followed by matter dominance and in the late time, dark energy is dominating the Universe (de-Silter), leading to the late time acceleration.

From the analysis of cosmic parameters with $x=0.1$ and $l=01.50$, we observed that the present value of $q$ is $-0.50$, indicating accelerated expansion of the Universe. With the present value of $\omega=-0.66$ indicates that the Universe is in the phase of Quintessence. Table (\ref{tab:parameters}) presents values of the state finder parameters and different conditions under which the Universe is in $\Lambda$CDM or Quintessence or Chaplygin Gas. The conditions for which the model is in $\Lambda$CDM are: always for $D$, $l=\sqrt{2}$ for $E$, $l=2$ for $F$.

Lastly, the validity of the model is checked with the available observation dataset viz. 77 Hubble data, 15 BAO data and 1048 pantheon data. These are done by taking $w=1.50$, $H_{0}=67.4$ km/s/Mpc, $\Omega_{m}=0.3153$ and $\Omega_{\Lambda}=0.6947$. For the first two datasets, we analyse the evolution of $H(z)$ and as predicted it is compatible with the $\Lambda$CDM model. For the pantheon dataset, we analyse the distance modulus function which is also align with that of the $\Lambda$CDM model. From the above discussion, we obtained, it can be concluded that our model is compatible with the observational evidence and make it a possible alternative to the cosmological model.


\begin{thebibliography}{100}
\bibitem{A.G. Riess} A.G. Riess et al., Astron. J., \textbf{116}, 1009 (1998).
\bibitem{S. Perlmutter} S. Perlmutter et al., Astrophys. J. \textbf{483}, 565 (1997).
\bibitem{S. Perlmutter M.S.} S. Perlmutter, M.S. Turner, M. White, Phys. Rev. Lett. \textbf{83}, 670 (1999).
\bibitem{D.J. Eisenstein} D.J. Eisenstein et al., Astrophys. J. \textbf{633}, 560 (2005).
\bibitem{L. Baudis} L. Baudis, J. Phys. G, \textbf{43(4)}, 044001 (2016).
\bibitem{S. Weinberg}S. Weinberg, Cosmology.
\bibitem{L. Pérénon} L. Pérénon, F. Piazza, C. Marinoni, et al.,J. Cosmol. Astropart. Phys., \textbf{2015(11)}, 029 (2015).
\bibitem{Planck2018} N. Aghanim, Planck Collaboration, et al., A$\&$A \textbf{641}, A6 (2020).
\bibitem{S. M. Carroll} S. M. Carroll, Living Rev. Rel., \textbf{4} (2001).
\bibitem{Carloni} S. Carloni et al., Class. Quantum Grav., \textbf{22}, 4839 (2005).
\bibitem{amendola} F. Fmendola and S. tsujikawa, Physics Letters B, \textbf{660(3)}, 125-132 (2008).
\bibitem{Felice} A. D. Felice, S. Tsujikawa, Physics Letters B, \textbf{675(1)}, 1-8 (2009). 
\bibitem{zang} Zhang, Y., Li, H., Gong, Y. et al., Eur.Phys.J.C, \textbf{72}, 2035 (2012).
\bibitem{Santos} S. Santos Da Costa et al., Class. Quantum Grav., \textbf{35}, 075013 (2018).
\bibitem{Surrendra A.} A. Samaddar, S. Surendra, Commun. Theor. Phys. \textbf{77(4)}, 045403 (2025). 
\bibitem{Sonia S.} S. Surendra Singh, Chingtham Sonia, Advances in High Energy Physics, \textbf{2020(1)}, 1805350 (2020).
\bibitem{Shivangi S.} Rathore, S., Singh, S.S., Sci Rep, \textbf{13}, 13980 (2023).
\bibitem{s. SS} Rathore, S., Singh, S.S., Muhammad, S. et al., Eur. Phys. J. C, \textbf{84}, 1108 (2024).
\bibitem{U. Debnath} U. Debnath, Int. J. Mod. Phys. A \textbf{35}, 2050203 (2020).
\bibitem{E J Copeland} E J Copeland, A R Liddle, and D Wands, Physical Review D, \textbf{57(8)},4686 (1998).
\bibitem{V. Sahni} V. Sahni, T.D. Saini, A,A.Starobinsky, et al., J. of Exper. and Theor. phys. Lett., \textbf{77}, 201-206 (2003).
\bibitem{J.K. Singh} J.K. Singh, H. Balhara, Shaily, P. Singh, Astronomy and Computing, \textbf{46}, 100795 (2024).
\bibitem{Amit s.} A. Samaddar, S. Surendra Singh, Phys. of the Dark Universe \text{47}, 101792 (2025).
\end{thebibliography}
  \end{document}